\documentclass[aps,pra,superscriptaddress,twocolumn]{revtex4}
\usepackage{amsmath,epsfig,mathrsfs,graphicx,amssymb}
\usepackage[T1]{fontenc}
\usepackage{t1enc}
\usepackage{hyperref}
\usepackage{yquant}
\useyquantlanguage{qasm}
\usepackage{physics}
\def\mb{\begin{pmatrix}}
\def\me{\end{pmatrix}}
\def\be#1\ee{\begin{equation}#1\end{equation}}
\newcommand{\ba}{\begin{eqnarray} }
\newcommand{\ea}{\end{eqnarray} }

\begin{document}

\title{Testing accuracy of qubit rotations on a public quantum computer}

\affiliation{Faculty of Physics, University of Warsaw, ul. Pasteura 5, PL02-093 Warsaw, Poland}
\affiliation{Systems Research Institute, Polish Academy of Sciences, 6 Newelska Street, PL01-447 Warsaw, Poland}
\affiliation{Center for Theoretical Physics, Polish Academy of Sciences, Al. Lotnik{\'o}w 32/46, PL02-668 Warsaw, Poland}

\author{Tomasz Bia{\l}ecki$^{1}$}
\author{Tomasz Rybotycki$^{2,3}$}
\author{Jakub Tworzyd{\l}o$^{1}$}
\author{Adam Bednorz$^{1}$}

\email{Adam.Bednorz@fuw.edu.pl}

\date{\today}

\begin{abstract}

We analyze the results of the test of $\pi/2$ qubit rotations  on the public quantum computer provided by IBM.
We measure a single qubit rotated by $\pi/2$ about a random axis, and we accumulate vast statistics of the results. 
The test performed on different devices shows systematic deviations from the theoretical predictions, 
which appear at the level $10^{-3}$. Some of the differences, beyond 5 standard deviations, cannot be explained by simple corrections due
to nonlinearities of pulse generations.
The magnitude of the deviation is comparable with the randomized benchmarking of the gate, 
but we additionally observe a pronounced parametric dependence. 
We discuss other possible reasons of the deviations, including states beyond the single-qubit space. 
The deviations have a similar structure for various devices used at different times, 
and so they can also serve as a diagnostic tool to eliminate imperfect gate implementations, and faithful description of the involved physical systems.

\end{abstract}

\maketitle

\section{Introduction}

Recent progress in the operation of quantum devices offered by IBM 
enables many researchers to perform quantum computations in a realistic setup \cite{alsina,alsina2,devitt,mizel,rundle,berta}. The fragility of the operating devices, user-defined actions, and readouts deserve constant diagnostic checks. The paradigm for operating these systems relies on the quantum description of few-level Hilbert space and the unitary evolution controlled by a programmed sequence of gates.
The devices and operations are not perfect in reality: the deviations come from the decoherence, the environment noise, inaccuracy of the gate parameters, and the presence of additional states. To diagnose realistic implementation of the ideal model, one can perform various control tests, where the outcome statistics reveal the nature of such deviations, their possible sources, and hints for countermeasures \cite{opt}.

In this paper, we propose to perform a simple experiment as a diagnostic test of the reliability of the quantum gates. 
In short, the test compares the outcome probability of the qubit in a specific state with the standard $\cos^2\theta$, with $\theta$ 
being an angle of axis of the $\pi/2$ rotation gate. 
Taking a list of angles, shuffled randomly, and repeating the test sufficiently many times, one can reveal potential deviations.
There exist other testing approaches based on 
the state dimension \cite{opti}. 
The specifics of our test are: minimalistic circuit complexity, single control parameter (angle $\theta$), robustness to many sources of noise. 
The test is also linear, being robust against drifts and calibration changes. Our test goes beyond the standard randomized benchmarking \cite{benchpap}, 
as we systematically monitor the dependence of the deviations on a control parameter. 

The public quantum computer, Quantum Experience by IBM, offers the possibility to perform such a test with sufficiently large statistics,.
We were able to run the experiment on several different devices, including a single-qubit one. 
The statistics we collected were sufficiently large to make confident conclusions. We found deviations at the relative level $10^{-3}$, 
and far beyond 5 standard deviations. Our observations show that the deviations are not accidental and  the corrections 
to the ideal model are necessary to explain them. We also tested nonlinearities of the waveform generators \cite{distor} as the possible cause and
 they only partially explain the data.
The remaining discrepancies are still beyond 5 standard deviations and their cause is still unknown. There may be subtle technical reasons, but
extraordinary models such as involving larger Hilbert space \cite{dimwitness} or more exotic concepts like interacting many copies \cite{plaga,abadp} should also be considered. 
We perform additional benchmarking tests to show that the deviations are independent of the inevitable decoherence caused by subsequent gates.

The paper is organized as follows. We start by describing the test of the $\pi/2$ rotation on a qubit, repeated $n$ times, then explain implementation on IBM cloud computing, 
next discuss the obtained results and their significance, including the analysis of $n=1,5$ and $n=1,5,9$ cases. 
We discuss the benchmark tests and close with the general summary. We present the calculation of model-based deviations in the Appendixes.

\begin{figure}[ht!]
	\centering
	\begin{tikzpicture}[scale=1.3]
		\begin{yquant*}
			init {$\ket 0$} q[0];
			box {$S$} q[0];
			box {$S_\theta$} q[0];
			[draw=none]
			box {$\dots$} q[0];
			box {$S_\theta$} q[0];
			measure q[0];
		\end{yquant*}
	\end{tikzpicture}
	\caption{The quantum circuit used to test the $\pi/2$ rotation. The initial state $\ket 0$ is rotated by the gate $S$, and the sequence of identical gates $S_\theta$ is applied before measurement.}\label{cir}
\end{figure}
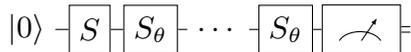
\begin{figure}
\includegraphics[scale=.6]{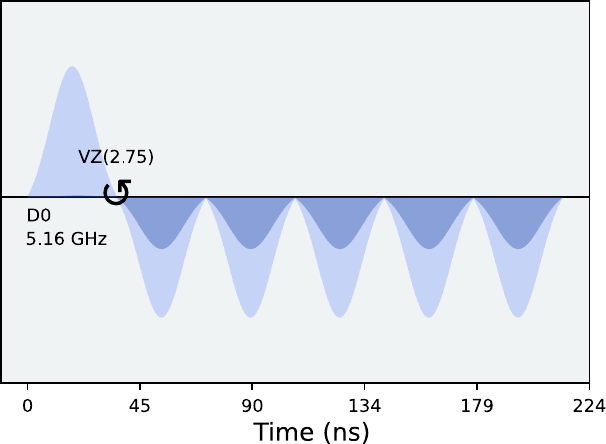}
\caption{The actual waveform of the pulse on IBM quantum computer (perth), first $S$, then $n=5$ gates $S_\theta$ for $\theta=7\pi/8\approx 2.75$. 
The discretization unit time is $dt=0.222$ns. 
Driving (level gap) frequency is denoted by $D0$. The light/dark shading corresponds to in-phase/out-of-phase amplitude component, respectively. The element $\mathrm{VZ}(2.75)$ is a zero-duration virtual gate, a part of native $S_\theta$.}
\label{wav}
\end{figure}

\section{Test of $\pi/2$ rotation on a qubit}

We use a minimal set of operations to prepare a parameter dependent linear combination of the 
ground state $\ket{0}$ and the excited state $\ket{1}$ in the Hilbert space of a qubit. The operations we use conform to native gates \cite{native} provided by IBM Quantum cloud computing.

 We assume  a $\theta$-dependent quantum operation (gate) is of a general form
\be
S_\theta=Z^\dag_\theta SZ_\theta \label{sthe},
\ee
with the angle-dependent rotation around $z$ axis on Bloch sphere
\be
Z_\theta=
\begin{pmatrix}
e^{-i\theta/2}&0\\
0&e^{i\theta/2}\end{pmatrix}\label{zzz}
\ee
written in the basis $|0\rangle$, $|1\rangle$.
The rotation $Z_\theta$ is virtual and is performed together with $S$ \cite{zgates} resulting in a single operation $S_\theta$.

We intend to use the native gate $S\equiv \sqrt{X}$ as a simplest choice. 
The operation $S_\theta$ can be applied $n$ times resulting in the total operation $S_\theta^nS$ 
acting on the initial ground state $|0\rangle$. The sequence of operations is depicted schematically in 
Fig.\ref{cir} with the details of pulse shapes shown in Fig.\ref{wav}.

We perform a dichotomic diagonal measurement  
$M=\alpha|0\rangle\langle 0|+\beta|1\rangle\langle 1|$ on the prepared state $\rho$, which gives the mutually exclusive outcomes $1,0$. 
A general form for the  probability of $1$, given some initial $\rho$ and some given $S$ is 
\be
p_{n\theta}=\mathrm{Tr}MS_\theta^n\rho S^{n\dag}_\theta
=A_n\sin \theta+B_n\cos \theta+ C_n\label{abc}
\ee
with some constants $A_n,B_n,C_n$, which are independent of $\theta$.

Of course, this prediction will no longer be valid if (a) the actual Hilbert state is larger, 
with e.g. an additional state $|2\rangle$, (b) $M$, $S$ or $\rho$ depend on $\theta$, (c) $M$ is not diagonal. 

Only such effects can lead to deviations from (\ref{abc}). For (c), only a second harmonics occurs.
We shall discuss these possibilities in detail in Sec. \ref{results}, focusing on potential perturbative corrections. 
The perfect operation corresponding to our particular choice is given by $M=|1\rangle\langle 1|$, $\rho=S|0\rangle\langle 0|S^\dag$ and  
$S\equiv RX(\frac{\pi}{2})\equiv \sqrt{X}$ . Note that $S$ is just a  $\pi/2$ rotation around $x$ axis on Bloch sphere
\be
S=\frac{1}{\sqrt{2}}\begin{pmatrix}
1&-i\\
-i&1\end{pmatrix}\label{smat}.
\ee
It is important that an ideal  $\pi/2$ rotation has eigenvalues $\pm 1$ and $1$.
We are going to test (A) the fit from (\ref{abc}) against measured outcome of the preparation sequence for a specific $n$ and (B) if $p_1-p_5=0$
and $p_1-2p_5+p_9=0$ for an arbitrary $\theta$.
The great advantage of these tests is the usage of only a single qubit, partial independence of unknown properties of 
the environment and quantum operations and universality -- applies to any two-level system. In practical implementation it is helpful 
to eliminate memory effects of its repetitions, by picking a random $\theta$ from a range uniformly covering any interval of length $2\pi$. 
The result of the test should not change if adding a definite number of operations $S$ at the end. 
Note only that in the ideal case an even number of $S$s after $Z_\theta$ gives the probability $1/2$ 
while for an odd one the probabilities of $0$ and $1$ get swapped every two $S$s.

\begin{figure}[ht!]
\includegraphics[scale=0.72]{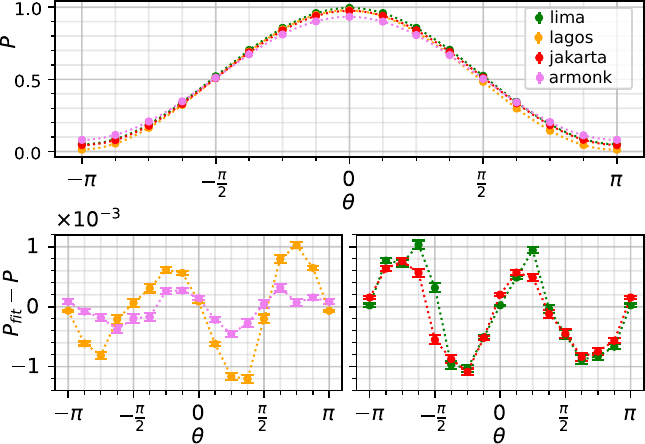}
\caption{ The results of the tests on IBM quantum devices for a single $S_\theta$ with 8192 shots per job with 56 circuits per angle per job and 100 jobs, except armonk with 4 circuits per angle and 1556 jobs. The probability of registering $1$ is fitted by least squares to (\ref{abc}) in the upper figure, while the lower figure presents the deviation from the fit. The errors are given by Bernoulli formula 
for the variance $\sqrt{p(1-p)}$ times the number of repetitions (jobs times shots times circuits per angle). On the vertical axis $P$ denotes the probability of $1$ while $P_{\mathrm{fit}}$ is the fit of $P$ to (\ref{abc}).}
\label{res}
\end{figure}

\begin{figure}[ht!]
\includegraphics[scale=0.72]{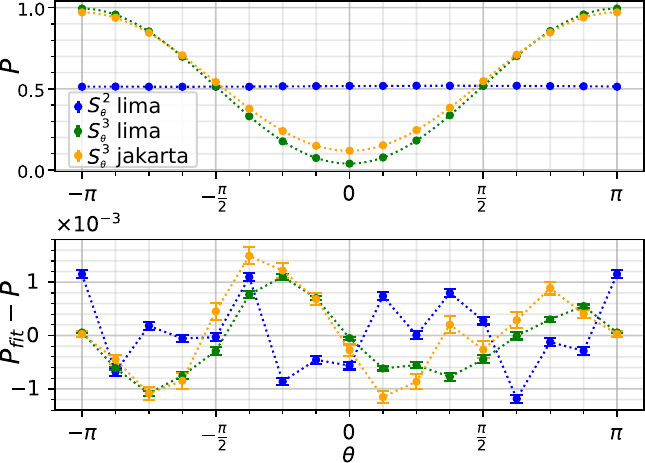}
\caption{The results of tests as in Fig. \ref{res} but with two and three $S_\theta$. Note that 
the ideal angle dependence for $n=3$ is reversed with respect to Fig. \ref{res}, as the two $S_\theta$ swap the states $|0\rangle$ and $|1\rangle$.}
\label{res2}
\end{figure}

\section{Implementation on the quantum computer}

The IBM Quantum Experience cloud computing offers several devices, working as a collection of qubits, which can be manipulated
by gates from a limited set -- either single qubit operations or twoqubit ones. Some gates can depend on a real parameter. The sequence of the gates
is user-defined. The provided interface allows some fine tuning, like delay of the gate, using barriers or performing additional resets. Physically the qubits are transmons \cite{transmon}, the artificial quantum states existing due to superconductivity and capacitance (interplay of Josephson effect and capacitive energy). In principle the transmon has more than two states but the gates' implementation is tailored to  limit the working space to two states.
The time of decoherence (mostly due to environmental interaction) is sufficiently long to perform a sequence of quantum operations and read out reliable results. 

The ground state $|0\rangle$ is the lowest energy eigenstate of the transmon, that can be additionally assured by a reset operation. Gates are time-scheduled microwave pulses prepared by waveform generators  and mixers (of length $30-70$ns with sampling at $0.222$ns) , taking into account the frequency equivalent to the energy difference between qubit levels \cite{qis} (about $4-5$Ghz).
The rotation $Z$ is not a real, but an instantaneous virtual gate $\mathrm{VZ}(\theta)$, added to the next gate \cite{zgates}. In particular the sequence of gates $Z_\theta$ and $S$
is realized by the native $S_\theta$. The readout is realized by coupling the qubit with a resonator whose frequency depends on the qubit state 
and measure the phase shift of the populated photons \cite{qis,read}.

To run the experiment one has to prepare a script controlling the jobs sent to the computer, lists of individual circuits describing the sequence of the gates
and possible parameters, the number of  shots, i.e. the number of repetitions of the list of circuits, limited by 8192, later extended to 20000, 32000 and 100000, depending on the device.
Each device has some limit on the number of circuits per job. We used lima, jakarta and bogota, as they offered 900 circuits per job running about 100 jobs
(later lima and jakarta reduced to 100 and 300, respectively).
We also used armonk which is the only single-qubit device, but it offers only 75 circuits per job. In this case, to obtain significant statistics, we had to
run more than 1500 jobs. Each circuit consisted of a sequence of gates $S$ and $S_\theta$ for $\theta=j\pi/8$, $j=-7,-6,...,7,8$ (16 even-spaced values) 
with additional resets at the beginning and after the readout. This eliminates effect of daily calibrations. The typical circuits is depicted in Fig. \ref{cir} with the actual pulse sequence shown in Fig. \ref{wav}. To avoid memory effects, we shuffled randomly the values of $\theta$ individually in each job.
We compared the outcome statistics, the measured rate of occurred value $1$, with the fit to (\ref{abc}).
We have also run the circuits consisting of two and three gates $S_\theta$ instead of one, to compare the possible deviations. For a benchmark, to estimate an error per gate, we used up to 70 $S_\theta$. We made our scripts and collected data publicly available \cite{zen}.

\section{Results}
\label{results}

We present the results in Fig. \ref{res}, \ref{res2}. In all runs, the prediction (\ref{abc}) is verified down to the level $10^{-3}$ of relative error.
However, the deviations of the order $10^{-3}$ are significant when compared to the predicted error (more than 5 variances).
The deviation is smallest on armonk -- the single-qubit device, but still significant. Note that
the execution time of $S$ or $S_\theta$ gates is $35.5$ns except on armonk, where it is $71.1$ns. As the similar results, exhibiting systematic $\theta$-dependent deviations, have been obtained on different devices
in different times (the data collection took one year, 2021-2022, while each run took from a few days to a few weeks) they deserve some physical explanation. The results from bogota are consistent, but indicate a noticeable phase shift, we discuss them separately Appendix \ref{apa}.
The natural reason  is that the actual performance of the gates can differ from the ideal expectations. 
For instance a nonlinearity of the waveform generator can modify the pulse in a $\theta$-dependent way \cite{distor}. 

We analyze below a non-ideal execution of the gates under 4 assumptions
\begin{enumerate}
\item The 2-dimensional Hilbert space of quantum system.
\item The identical subsequent $S_\theta$ gates.
\item Decoherence independent on $\theta$-parameter.
\item The small deviations, i.e. dominating the first order correction.
\end{enumerate}
The standard realization of the gate $S$ including generic deviations in the actual pulse can be  described as
\be
S_\theta=\mathcal T\exp\int_0^{\pi/2} (e^{i\theta}(1+\epsilon)\ket 0 \bra 1+\mathrm{h.c.})d\phi/2i \label{gat}
\ee
for some complex $\epsilon(\theta,\phi)=\epsilon_r+i\epsilon_i$,
where $\mathcal T$ denotes chronological product in the Taylor expansion of the exponential of the integral with respect to $\phi$ corresponding to 
dimensionless gate operation time. The gate is ideal for $\epsilon=0$. We consider only off-diagonal corrections because the pulse only modulates the driving 
frequency of the transition between levels.
We find the first order correction to the probability (\ref{abc}) for $n$ gates $S_\theta$ in the form  (see Appendix \ref{apb})
\ba
&&\delta p_{n\theta}=\int_0^{\pi/2} d\phi\times\label{pcor}\\
&&\left\{
\begin{array}{ll}
-\sin\theta \sin\phi\;\epsilon_i/2&\mbox{ for }n\equiv 1\:\mathrm{mod}\:4\\
\sin\theta(\cos\phi-\sin\phi)\epsilon_i/2&\\
-\cos\theta\;\epsilon_rn/2&\mbox{ for }n\equiv 2\:\mathrm{mod}\:4\\
\sin\theta\cos\phi\;\epsilon_i/2&\mbox{ for }n\equiv 3\:\mathrm{mod}\:4\\
\cos\theta\;\epsilon_rn/2&\mbox{ for }n\equiv 0\:\mathrm{mod}\:4
\end{array}
\right.\nonumber
\ea
At first sight, the model above
could in principle explain the deviations, because most of deviations cross $0$ at $\theta=0,\pi$ (taking into account a general shift of the angle on bogota) and 
the deviations for a single and three $S_\theta$ sum approximately to zero if the symmetry $\epsilon(\phi)=\epsilon(\pi/2-\phi)$ is assumed. 
Nevertheless, this must be  confirmed by tracing down to the actual pulse formation. 
The results for $S^2_\theta$ presented in Fig. \ref{res2} are not fully compatible
with this model at $\theta=+\pi/2$ but it may be result of assumption of identical subsequent gates which may be not fully realized.

\begin{figure}[ht!]
\includegraphics[scale=0.68]{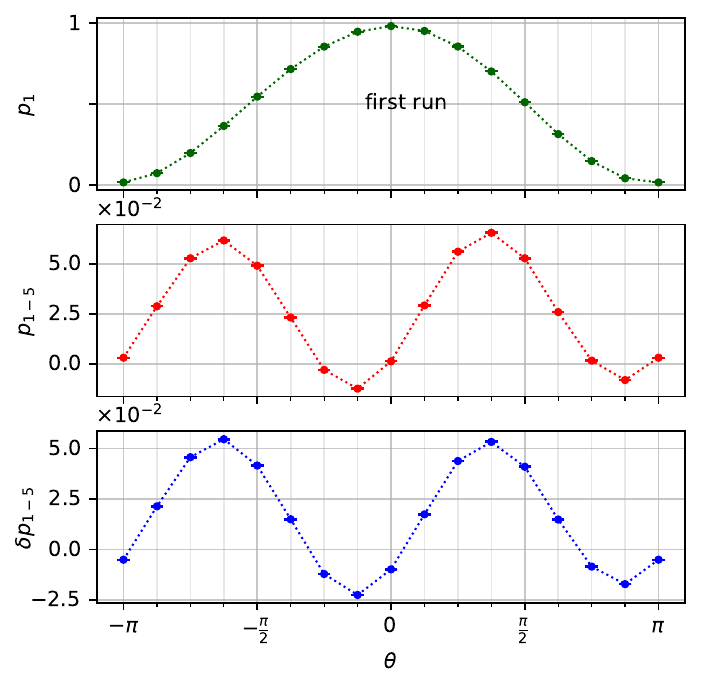}\\
\includegraphics[scale=0.68]{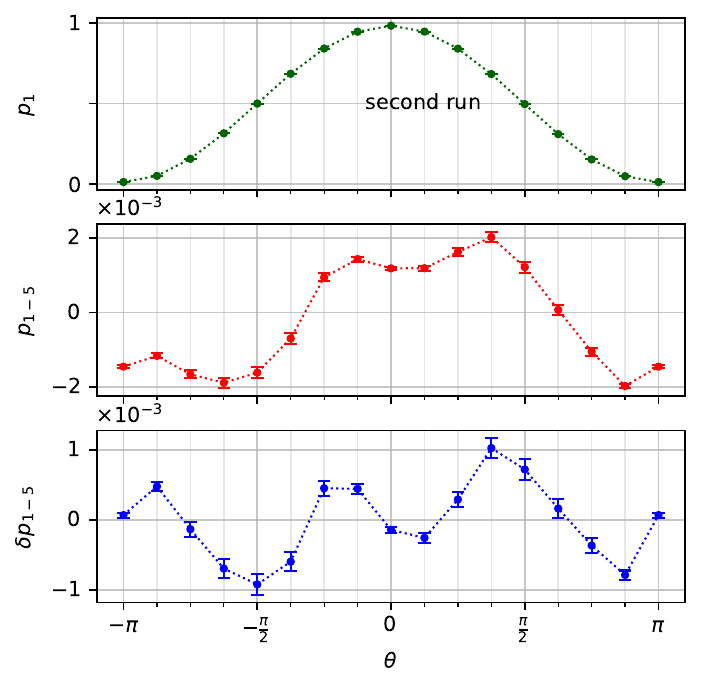}\\
\caption{Comparison between $1$ and $5$ gates, $p_{1-5}=p_1-p_5$ with the fit to (\ref{abc}) subtracted in $\delta p$, for lagos qubit $0$, using  32000 shots each and 9 repetitions for each case angle for 43 jobs in
the first run in February  and 80 jobs in the second run in May 2022,
each data point is a result of $43\cdot 32000\cdot 9$ or $80\cdot 32000\cdot 9$ runs.
Note the very large reversal deviation in the first run, reversed between $1$ and $5$.}
\label{reslag15fs}
\end{figure}

\begin{figure}[ht!]
\includegraphics[scale=0.73]{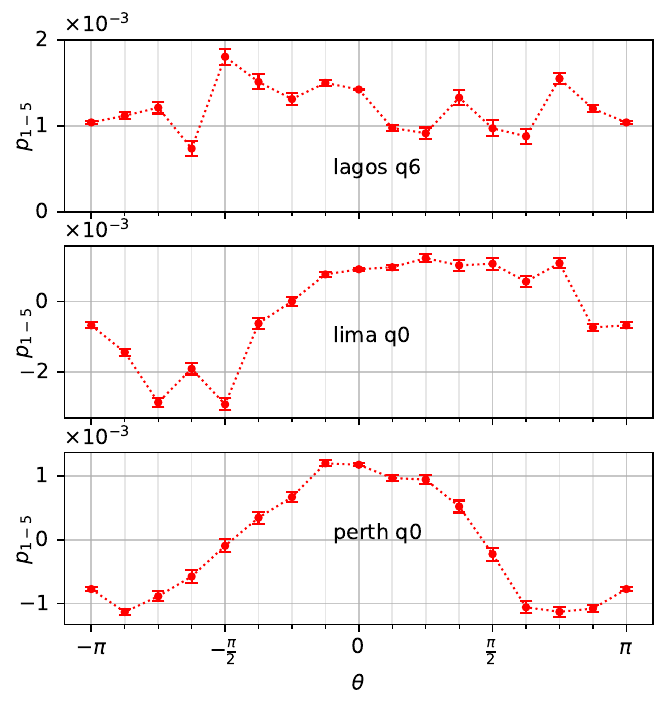}
\caption{Comparison between $1$ and $5$ gates, $p_1-p_5$, for lagos qubit $6$, with 195 jobs with 32000 shots each and 9 repetitions for each case angle,
(top), lima qubit 0 with 290 jobs, 20000 shots, 3 repetitions (middle), and perth qubit $0$, with 53 jobs, 100000 shots each and 9 repetitions (bottom).}
\label{rest15}
\end{figure}

\begin{figure}[ht!]
\includegraphics[scale=0.70]{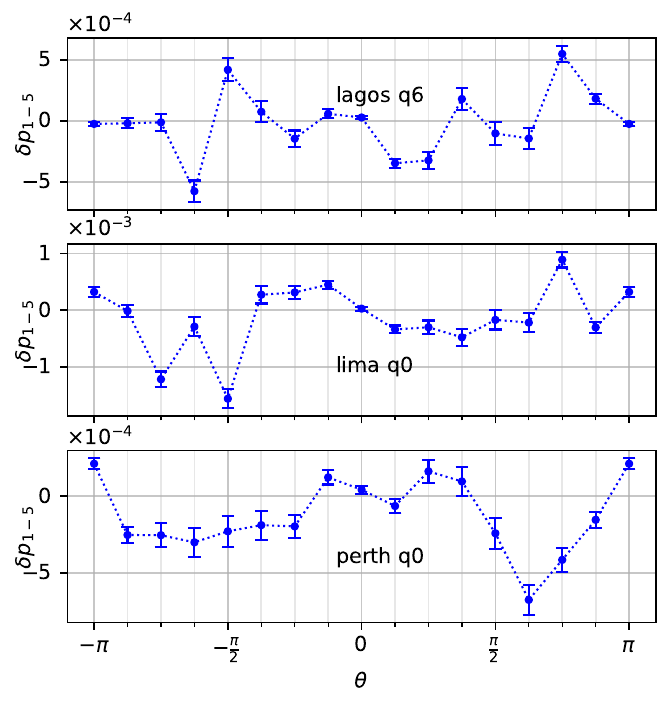}
\caption{Results for $1$ and $5$ gates, as in Fig. \ref{rest15}, after removing the fit to (\ref{abc})}
\label{rest15f}
\end{figure}

In order to fully test the model, we have re-run the tests to compare the results from $1$ and $5$ gates $S_\theta$ which should give identical deviations,
according to the model, i.e. $p_1-p_5=0$ in the first order. We have performed the tests on lagos (qubits 0 and 6), lima (qubit 0) and perth (qubit 0), running with shuffled angles to avoid
memory effects. The results show extraordinary deviations for lagos qubit $0$ in the first run in February 2022 
(however, we found
such large deviation already in November 2021 in the benchmark test), but repeating the test in May 2022 gave much smaller deviations, see Fig. \ref{reslag15fs}. 
The deviation from the first run is large, $\sim 10^{-2}$, and
gets inverted between $1$ and $5$ gates. We have additionally checked that the inversion took place each $4$ gates in the benchmark test.
Here the reason must  have been completely different, e.g. a considerable technical or fundamental problem. Such large deviation can be explained be 
an enlarged Hilbert space, including multiplication of quantum states, in analogy to many-copies idea \cite{plaga, abadp}, but this needs
further analysis to confirm or rule out. Smaller deviations in the second run indicate that they may depend on calibration, which is applied to the qubit daily,
although they stayed on the same level during 6 months, ruling out effect of an incidental calibration.
The qubit 6 from lagos and qubits 0 from lima and perth give also much smaller deviations, 
Fig. \ref{rest15} although still they do not match for $1$ and $5$ gates, small but still nonzero (beyond 5 variances) differences.

\begin{figure}[ht!]
\includegraphics[scale=0.70]{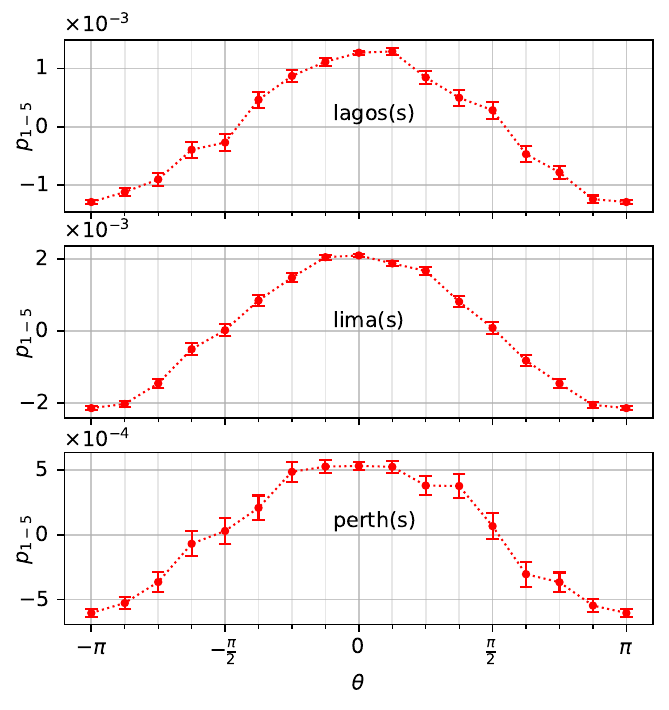}
\caption{Simulation of results of $p_1-p_5$ for $1$ and $5$ gates, as in Fig. \ref{rest15} taking noise models from: lagos with 80 jobs with 32000 shots each and 9 repetitions for each case angle,
(top), lima with 290 jobs, 20000 shots, 3 repetitions (middle), and perth  with 53 jobs, 100000 shots each and 9 repetitions (bottom).}
\label{sim}
\end{figure}

\begin{figure}[ht!]
\includegraphics[scale=0.70]{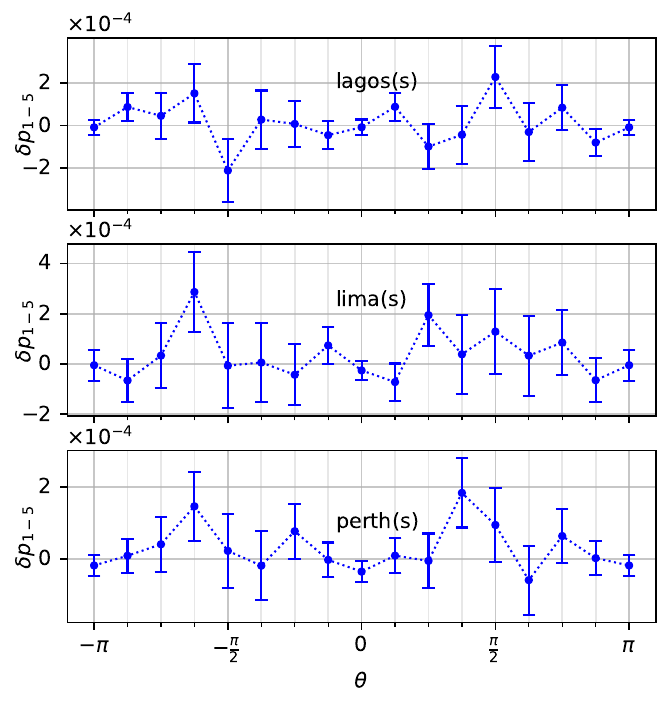}
\caption{Simulation of results for $1$ and $5$ gates, as in Fig. \ref{sim}, after removing the fit to (\ref{abc})}
\label{simf}
\end{figure}

\begin{figure}[ht!]
\includegraphics[scale=0.70]{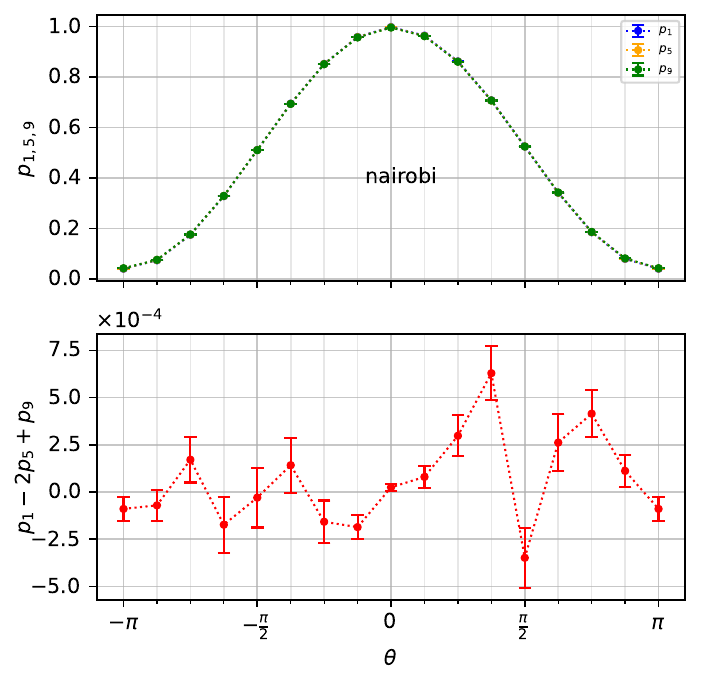}
\caption{Comparison between $1$,$5$ and $9$ gates, $p_1-2p_5+p_9$, for nairobi qubit $0$, with 100 jobs with  100000 shots each and 6 repetitions.
The actual probabilities almost overlap each other but the combination gives a deviation over 4 standard deviations. }
\label{nai159}
\end{figure}

\begin{figure}[ht!]
\includegraphics[scale=0.70]{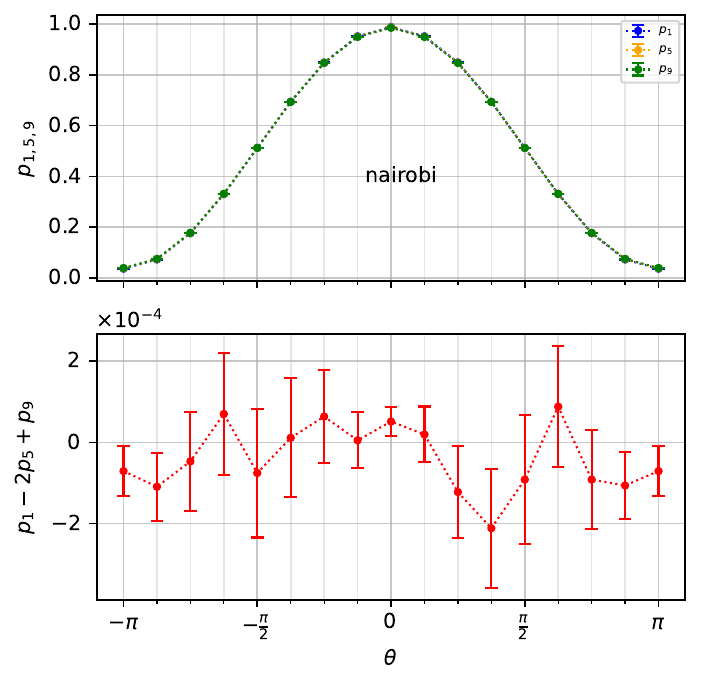}
\caption{Simulation of $1$,$5$ and $9$ gates, testing $p_1-2p_5+p_9$, using the noise model from nairobi, with the same number of jobs, 
shots and repetitions as in Fig. \ref{nai159}.}
\label{nai159sim}
\end{figure}

Standard models of decoherence do not depend on gate parameters ($\theta$ in our work). 
In Appendix \ref{apc} we present an analytical argument for such a model, including readout error, relaxation, 
depolarization and phase damping \cite{nielsen}, to show that $p_5\simeq p_1$ still holds up to first-order corrections. 
More importantly, we performed simulations on IBM using noise models from lagos, perth and lima, 
to show a very good agreement with (\ref{abc}). Within statistical errors, as illustrated in Fig. \ref{sim} and \ref{simf}, 
the results for one $S_\theta$ and five $S_\theta^5$ gates agree well with one another and with the fitting formula. 
The magnitude of the statistical errors compares favorably with the real device error estimates, such as in Fig.\ref{rest15}. 
Finally, we checked if $p_1-2p_5+p_9=0$ on nairobi (January-Ferbuary 2023) which should hold up to second order deviations, see Appendix \ref{apc}.
The result is beyond 4 standard deviations, see Fig. \ref{nai159}, while the simultation does not show a deviation, see Fig. \ref{nai159sim}.
We conclude that the real device deviation from a test value, which we observe, goes beyond the standard models of noise, and the known
amount of leakage \cite{leak}.

A higher state \cite{dimwitness} as an alternative explanation seems unlikely due to different transition frequencies and the fact that it is a second order correction (see the analysis in the Appendix \ref{apd}). A simple in-phase/quadrature ($I/Q$) imbalance \cite{zgates} cannot explain the dominating 2nd harmonic in the deviations, as it would give
only 3rd harmonics (see Appendix \ref{ape}). An even more complicated description, like considerable extension of the Hilbert space \cite{plaga,abadp} would be the last option.

\begin{figure}
\includegraphics[scale=.8]{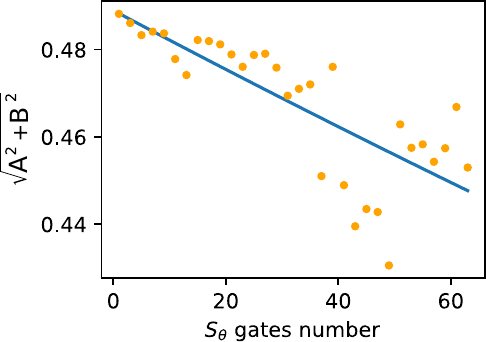}
\caption{The decay of the amplitude $\sqrt{A^2+B^2}$ with the number $n$ of gates ($n$ is odd). By least squares fit to the formula \ref{fit} we estimated $r\simeq 7\cdot 10^{-4}$.
The test has been run on lagos with $n=63$ jobs, each corresponding to a subsequent number of $S_\theta$, 8192 shots per job and 56 circuits per angle.}
\label{ben}
\end{figure}

\begin{figure}[ht!]
\includegraphics[scale=0.6]{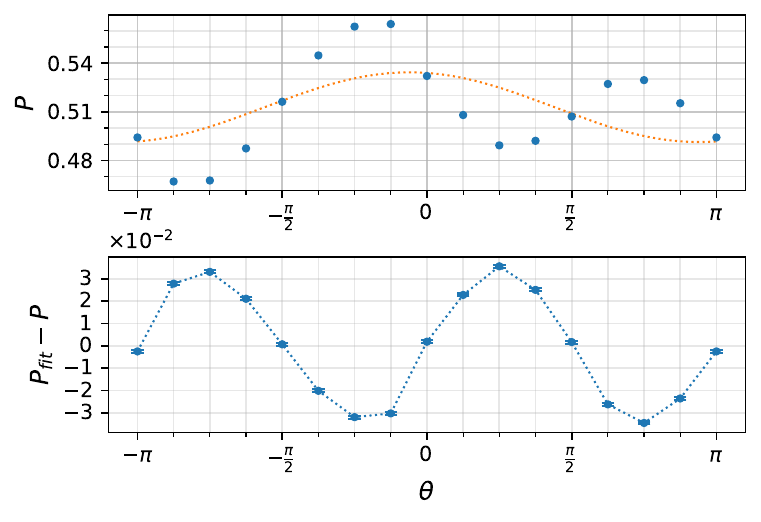}
\includegraphics[scale=0.6]{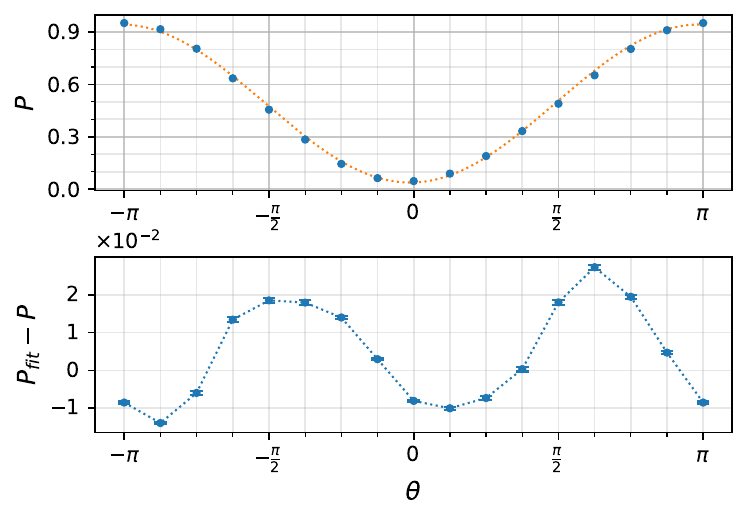}
\caption{The fit to (\ref{abc}) and deviations after $n=62$ (upper) and $n=63$ (lower) $S_\theta$, in the same experiment on lagos as in Fig. \ref{ben}. }
\label{abc63}
\end{figure}

\section{Benchmark}

In addition to the above tests, we have checked on bogota how the amplitudes of the fit, i.e. coefficients $A$ and $B$ decrease with an increased number $n$ of $S_\theta$ gates, in analogy to the standard benchmark tests \cite{bench,deep}. The decay of $A$ and $B$ over the number of gates corresponds to the decoherence induced by gates and environment. For an odd number $n$ the signs of $A$ and $B$ alternate every two $S_\theta$ gates.  We estimate the error-per gate $r$ with a fit to the formula
\be
\sqrt{A^2+B^2}=(1-r)^n D. \label{fit}
\ee
(For even $n$ the ideal expectation is $A=B=0$, so we don't include these data.)

We found that the error gets accumulated as confirmed by checking the fit after $n=62$  and $n=63$ $S_\theta$, see Fig.\ref{abc63} and cannot be explained by the first order deviations (\ref{pcor}) meaning that other effect may be comparable. 
Nevertheless, the normalized error per gate, $r\sim 7\cdot 10^{-4}$, estimated from the data presented in Fig. \ref{ben}, 
remains smaller than our deviations. We conclude that they must have a different origin.  
In addition, if the error is caused by leakage to other states then it is unlikely that it will cause $\theta$-dependent deviation of the same order 
(at least second order, see Appendix \ref{apc}).

\section{Discussion}

We have observed the deviation from the ideal $\pi/2$ rotation on several devices available at IBM Quantum Experience.
The deviations are significant, systematic, and with the amplitude exceeding 5 standard deviations. They exist in a similar form on different devices, tested over a long period of time. The deviations are smaller, but persistent
for a single-qubit armonk. The benchmark test rules out the possibility of accumulation of decoherence error by many identical gates. Also the angle-dependent contribution from higher states should remain negligible in the lowest order.
The most likely solution, the imperfection of $I/Q$-mixers and waveform generators, close to $\sin 2\theta$
(except bogota), fails to reproduce equal deviations in the case of $1$ and $5$ gates $S_\theta$, so it is at best insufficient.
The additional test of $1$, $5$, and $9$ gates gives still the deviation beyond 4 variances, which deserves  confirmation in a
larger statistics, colleced within a realtively short time (in our case below 2 months).
Nevertheless, the systematic occurrence of the deviation should serve as a diagnostic test to enhance calibration of the gates,
and find the correct description of the qubit. Due to our assumptions on the identical pulses and restricted Hilbert space,
we cannot claim the deviations to be a signature of fundamental problems with the description of transmon qubit.
However, we believe that the robustness of such tests will motivate to further exploration of qubits diagnostics.

One can continue the tests using OpenPulse API \cite{openpulse}, which allow to fine-tune the gates, or use more complicated tests
to reveal the relevant dimensionality of transmon Hilbert space.
In any case, we believe
that further improvement of  quantum computers from IBM or possible other public providers will allow even more stringent test of quantum predictions
in the case of low-level systems.

\section*{Acknowledgements}
We acknowledge use of the IBM Quantum Experience for this work. The views expressed are those of the authors and do not reflect the official policy or position of IBM or the IBM Quantum Experience team.)
\appendix
\section{Results from bogota}
\label{apa}

\begin{figure}[ht!]
\includegraphics[scale=0.7]{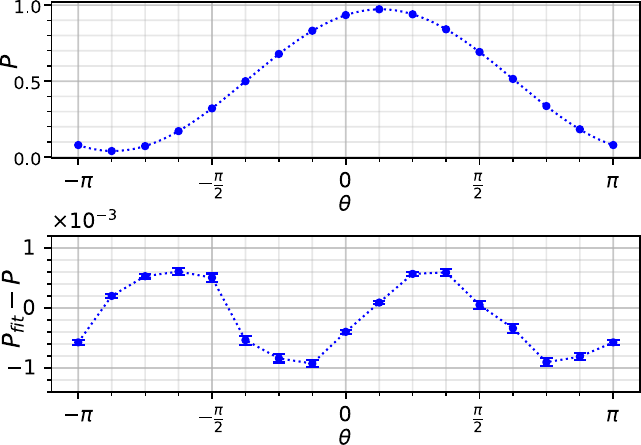}
\caption{ The results of the tests on bogota, with with 100 jobs, 8192 shots per job and 56 circuits per angle. Notation as in Fig. \ref{res}}
\label{resb}
\end{figure}

The results from the experiment on bogota, Fig. \ref{resb} revealed a global phase shift. We are not aware of its reason which can be either
due to a wrongly programmed gate or a transpiling error when the script is translated to physical instruction to be performed sequentially on the gates. 
Nevertheless the deviations are consistent with other devices, if the shift is taken into account. 
We stress that we used exactly the same script as for the other devices and bogota was neither the first not the last of devices to test.

\section{First-order deviations from ideal gate models}
\label{apb}
For a general gate (\ref{gat}) we define

\ba
&&U_\theta(\phi)=\exp \phi(e^{i\theta}\ket 0\bra 1+\mathrm{h.c})/2i\nonumber\\
&&=\begin{pmatrix}
\cos\phi/2& -ie^{i\theta}\sin\phi/2 \\
-ie^{-i\theta}\sin \phi/2&\cos\phi/2
\end{pmatrix}.
\ea
The correction in the first order of $\epsilon$ to the gate operation (\ref{gat}) reads then
\ba
&&\delta(S^n_\theta)=\nonumber\\
&&S^n_\theta\int_0^{n\pi/2}U^\dag_\theta(\phi)\begin{pmatrix}
0& -i\epsilon e^{i\theta}\\
-i\epsilon^\ast e^{-i\theta}&0\end{pmatrix} U_\theta(\phi)d\phi/2\nonumber\\
&&
=S^n_\theta\int_0^{n\pi/2} H'_\theta(\phi)d\phi/2i\label{des}
\ea
with
\be
H'_\theta(\phi)=\begin{pmatrix}
\epsilon_i\sin \phi&(\epsilon_r+i\epsilon_i\cos \phi)e^{i\theta}\\
(\epsilon_r-i\epsilon_i\cos \phi)e^{-i\theta}& -\epsilon_i \sin  \phi
\end{pmatrix}
\ee
assuming $\epsilon(\theta,\phi+\pi/2)=\epsilon(\theta,\phi)$ (periodicity in $\phi$ with respect to $\pi/2$ as the gates are identical).
The $\theta$-dependent correction to the probability $p_\theta=|\bra 1 S^n_\theta S\ket 0|^2$ reads
\be
\delta p_\theta=-2\mathrm{Re} \bra 0 S^\dag S^{\dag n}_\theta\ket 0 \bra 0 \delta(S^n_\theta)S\ket 0.\label{dpt}
\ee
Using (\ref{smat}) we calculate
\be
\rho=S\ket 0\bra 0 S^\dag=\begin{pmatrix}
1& i\\
-i& 1\end{pmatrix}/2.\label{r1s}
\ee
Denoting
\be
\rho_{n\theta}=S^{\dag n}_\theta MS^n_\theta,\label{stn}
\ee
for $M=\ket 0 \bra 0 $
we get  explicitly from (\ref{sthe}), (\ref{zzz}) and (\ref{smat})
\ba
&&\rho_{0\theta}=\begin{pmatrix}
1&0\\
0&0\end{pmatrix},\nonumber\\
&&\rho_{1\theta}=\begin{pmatrix}
1&-ie^{i\theta}\\
ie^{-i\theta}&1\end{pmatrix}/2,\nonumber\\
&&\rho_{2\theta}=\begin{pmatrix}
0&0\\
0&1\end{pmatrix},\\
&&\rho_{3\theta}=\begin{pmatrix}
1&ie^{i\theta}\\
-ie^{-i\theta}&1\end{pmatrix}/2,\nonumber
\ea
with $\rho_{n+4,\theta}=\rho_{n,\theta}$.
Substituting (\ref{des}), (\ref{r1s}) and (\ref{stn}) into (\ref{dpt}) we can write the deviation of (\ref{abc})
\be
\delta p_\theta=\int_0^{n\pi/2}d\phi\mathrm{Re}\mathrm{Tr}\:i\rho_1\rho_{n\theta}H'_\theta(\phi)
\ee
Since $\rho_1$, $\rho_{n\theta}$, $H'_\theta(\phi)$ are Hermitian, we get
\be
\delta p_\theta=\int_0^{n\pi/2}d\phi\mathrm{Re}\mathrm{Tr}\:i\rho_{n\theta}[H'_\theta(\phi),\rho_1]/2
\ee
with the commutator
\ba
&&i[H'_\theta(\phi),\rho_1]=\\
&&\begin{pmatrix}
\epsilon_r\cos\theta-\epsilon_i\cos \phi\sin\theta&-\epsilon_i\sin \phi\\
-\epsilon_i \sin  \phi&-\epsilon_r\cos\theta+\epsilon_i\cos \phi\sin\theta
\end{pmatrix}\nonumber
\ea
which allows to derive the final result (\ref{pcor}).

\section{Parameter-independent decoherence}
\label{apc}

The following reasoning shows that first order correction to the gate 
channels keeps equal deviations $\delta p_1\simeq \delta p_5$, regardless 
the way it is performed, only  but the final effect is taken into account.
Introducing standard $2\times 2$ Pauli matrices $\sigma_i$, $i=1,2,3$ and $\sigma_0$ denoting identity, every qubit state can be written
$
\rho=\boldsymbol n\cdot\boldsymbol \sigma/2
$,
with  $n_1^2+n_2^2+n_3^2\leq 1$, $n_0=1$ (equality for pure states).
Each quantum channel $\check{R}\rho$ is equivalent to
$R\cdot\boldsymbol{n}$ with some $4\times 4$ matrix $R$, whose first row reads $1,0,0,0$.  
In the unitary case $R$ contains a rotation matrix in the subspace $i=1,2,3$.
For the ideal gate $S_\theta$, acting in the sense of a quantum channel on density matrices, it is a $\pi/2$ rotation with eigenvalues $1$, $\pm i$. 

The measurement probability reads
$
p_k=\mathrm{Tr}MS^k_\theta\rho=\boldsymbol m\cdot S^k_\theta \boldsymbol n
$
with $M=\boldsymbol m\cdot\boldsymbol \sigma$. For a simple, diagonal  measurement we specify $m_1=m_2=0$, while we keep a general initial state $\boldsymbol n$.

We use a polar decomposition of the gate matrix $S_\theta=U_\theta V  D V^{-1} U^{-1}_\theta$, where 
\begin{align}
&U_\theta=\begin{pmatrix}
1&0&0&0\\
0&\cos\theta&-\sin\theta&0\\
0&\sin\theta&\cos\theta&0\\
0&0&0&1\end{pmatrix},\:V=\begin{pmatrix}
1&0&0&0\\
0&1&0&0\\
0&0&1&1\\
0&0&i&-i\end{pmatrix},\nonumber\\
&
D_\theta=\begin{pmatrix}
1&0&0&0\\
0&1&0&0\\
0&0&i&0\\
0&0&0&-i\end{pmatrix}.\label{polar}
\end{align}

The first order contribution to the difference of deviations from the ideal case comes only from corrections of eigenvalues of $S_\theta$, i.e.
\be
p_5-p_1=4\boldsymbol m\cdot U_\theta V  \delta D_\theta V^{-1} U^{-1}_\theta\boldsymbol n,
\ee
where
\be
\delta D_\theta=\begin{pmatrix}
0&0&0&0&\\
0&\eta_\theta&0&0\\
0&0&\epsilon_\theta&0&\\
0&0&0&\epsilon^\ast_\theta\end{pmatrix}
\ee
contains empirical parameters $\eta_\theta\in \mathbb{R}$, $\epsilon_\theta\in\mathbb{C}$, describing general linear deviations of eigenvalues of $S_\theta$.

Assuming preparation of a ground state $|0\rangle$ initially rotated by $\pi/2$ around $x$-axis, i.e. 
\be n_1=n_3=0\label{mmnn}
\ee
we get
$
p_1-p_5=4m_3n_2\mathrm{Im}\epsilon_\theta\cos\theta
$.
There is no difference between $p_1$ and $p_5$ for purely unitary evolution, in linear order in perturbation. 
As long as $\xi$ and $\eta$ remains $\theta$-independent, $p_1-p_5$ remains a combination of one, 
$\cos\theta$, and $\sin\theta$ as in (\ref{abc}), even for initial $\boldsymbol n$ deviating from the ideal case (\ref{mmnn}).

Furthermore, a combination of 1, 5 and 9 gates gives 
\be
p_1-2p_5=p_9=4\boldsymbol m\cdot U_\theta V  (\delta D_\theta)^2 V^{-1} U^{-1}_\theta\boldsymbol n,
\ee
which makes is of the second order in $\epsilon$.

A dissipative part can be derived from generic Lindblad equation
\be
\partial_t\rho=i[\rho,H(t)]/
\hbar+\sum_m(L_m\rho L^\dag_m-\{L^\dag_m L_m,\rho\}/2)
\ee
It covers depolarization, phase damping, and relaxation processes \cite{nielsen}, 
which can be described by $L=\lambda\sigma_3$ or $L=\lambda\sigma_\pm$ with $2\sigma_\pm=\sigma_1\pm i \sigma_2$.
For all such combinations one can write
$
\partial_t\boldsymbol n=H\boldsymbol n+L\boldsymbol n
$
where
\be
L=\begin{pmatrix}
0&0&0&0\\
0&A&0&0\\
0&0&A&0\\
B&0&0&C\end{pmatrix}
\ee
with some empirical constants $A$, $B$, $C$, while $H$
is an infinitesimal rotation.
During the gate operation, in the first approximation, one can simply rotate the $1,2$ subspace basis by $U_\theta$, same as in (\ref{polar}),
$H=U_\theta E U_\theta^{-1}$ with some  $\theta$-independent operation $E$. As $L$ commutes with $U_\theta$,
 the corrections to eigenvalues of $S_\theta$ will be then also independent of $\theta$. As we mentioned earlier, in this case the difference $\delta p_1-\delta p_2$ is incorporated in the fit (\ref{abc}).

\section{Corrections from higher states}
\label{apd}

We denote the basis states $|n\rangle$, $n=0,1,2,...$ and set $\hbar=1$.
The generic Hamiltonian
\be
H=\sum_n \omega_n|n\rangle\langle n|+2\cos(\omega t-\theta)\hat{V}(t)
\ee
consists of its own energy levels (first term) and the external influence given by frequency $\omega$, phase shift $\theta$ and the time-dependent pulse $\hat{V}$ (the second term). In principle free parameters $\omega,\theta$ and $\hat{V}(t)$ can model a completely arbitrary evolution.
However, the practical realization of gates implies separation of $\hat{V}$ into the ideal part and deviations. In this way, we can estimate deviations by perturbative analysis.
In addition, we set $\omega_0=0$, $\omega_1=\omega$ (resonance), $\omega_2=2\omega+\omega'$ (anharmonicity, i.e. $\omega'\ll\omega$, here about $300$Mhz). We restrict to the states $0,1,2$ which should contribute to the largest corrections.
Rotation and phase can be incorporated to the definition of states, $|n\rangle\to e^{-in(\theta+\omega t)}|n\rangle$.
In the new basis
\begin{widetext}
\be
H'=
\mb
2\cos(\omega t-\theta)V_{00}&(1+e^{-2i(\theta+\omega t)}) V_{01}&(e^{-i(\theta+\omega t)}+e^{-3i(\theta+\omega t)}) V_{02}\\
(1+e^{2i(\omega t+\theta)})V_{10}&2\cos(\omega t+\theta)V_{11}&
(1+e^{-2i(\theta+\omega t)}) V_{12}\\
(e^{-i(\theta+\omega t)}+e^{3i(\omega t+\theta)}) V_{20}&
(1+e^{2i(\omega t+\theta)}) V_{21}&
2\cos(\omega t+\theta)V_{22}+\omega'
\me
\ee
We split  $H'=H_{RWA}+\Delta H$ into the Rotating Wave Approximation (RWA) part
\be
H_{RWA}=
\mb
0&V_{01}&0\\
V_{10}&0&
 V_{12}\\
0&
V_{21}&
\omega'
\me
\ee
and correction
\be
\Delta H=\mb
2\cos(\omega t+\theta)V_{00}&e^{-2i(\theta+\omega t)} V_{01}&(e^{-i(\theta+\omega t)}+e^{-3i(\theta+\omega t)}) V_{02}\\
e^{2i(\omega t+\theta)}V_{10}&2\cos(\omega t+\theta)V_{11}&
e^{-2i(\theta+\omega t)} V_{12}\\
(e^{-i(\theta+\omega t)}+e^{3i(\omega t+\theta)}) V_{20}&
e^{2i(\omega t+\theta)} V_{21}&
2\cos(\omega t+\theta)V_{22}
\me
\ee
\end{widetext}
Evolution due to RWA reads
\be
U(t)=\mathcal T\exp\int_{-\infty}^t  H_{RWA}(t')dt'/i
\ee
where $\mathcal T$ means chronological product in Taylor expansion.
The full rotation is $U(+\infty)$. Only the state $|2\rangle$
contains 2nd harmonics $e^{\pm 2i\theta}$ after restoring original phases.

The 1st order correction to $U$ reads
\be
\Delta U=U(+\infty)\int dt U^\dag(t)\Delta H(t)U(t)/i
\ee
All terms in $\Delta H$ with $\theta$, contain $e^{i\omega t}$, too, which exponentially damps slow-varying expressions, e.g.
\be
\int e^{i\omega t}e^{-t^2/2\tau^2}dt\sim e^{-\omega^2\tau^2/2}
\ee
The 2nd order correction reads
\ba
&&\Delta^2 U=-U(+\infty)\times\\
&&\int dt U^\dag(t)\Delta H(t)U(t)\int^t dt'U^\dag(t')\Delta H(t')U(t').\nonumber
\ea
Most of components get damped exponentially, except when $\Delta H(t)$ contains $e^{ik\omega t}$
and $\Delta H(t')$ contains $e^{-ik\omega t}$, $k=1,2,3$, but even then $k\theta$ gets canceled
Therefore the nonnegligible part of $\Delta^2 U$ is independent of $\theta$ (compare with Bloch-Siegert shift \cite{blochs}), and
 can be observed as small heating, givng leakage at the level $10^{-5}$ \cite{leak}.
Due to a very short sampling time, $dt=0.222$ns, stroboscopic corrections to RWA \cite{rwa} can be neglected, too

\section{$I/Q$ imbalance}
\label{ape}

In reality the amplitude is a mixture of in-phase and out-of phase (quadrature) components, which may 
show some imbalance \cite{zgates} when mixing with local oscillator of frequency $\omega$,
\ba
&&V(t)=V_I(t)\cos(\omega t)+\\
&&V_Q(t)((1+\varepsilon_1)\sin(\omega t)+\varepsilon_2\cos(\omega t))\nonumber
\ea
with
\be
V_I(t)=\Omega(t)\cos\theta,\:V_Q(t)=\Omega(t)\sin\theta.
\ee
The minimal model of $I/Q$ imbalance in the basis $|0\rangle$, $|1\rangle$ is
\ba
&&H=\\
&&\mb
0&\Omega(t)e^{i\omega t}(e^{i\theta}+\varepsilon e^{-i\theta})\\
\Omega(t)e^{-i\omega t}(e^{-i\theta}+\varepsilon^\ast e^{i\theta})&\omega
\me\nonumber
\ea
where real $\Omega(t)$ determined the pulse shape and $\varepsilon=(-\varepsilon_1-i\varepsilon_2)/2$ is a small constant dimensionless complex number determining $I/Q$ imbalance.
With
\be
R_t=RR_{\omega}=\mb 
1&0\\
0&e^{i\theta}
\me\mb 
1&0\\
0&e^{-i\omega t}
\me
\ee
we remove the rotation of the Hamiltonian
\ba
&&R_t^\dag H R_t-R_t^\dag\partial_t R_t/i=H'=\nonumber\\
&&\mb
0&\Omega(t)(1+\varepsilon e^{-2i\theta})\\
\Omega(t)(1+\varepsilon^\ast e^{2i\theta})&0
\me
\ea
In the following we exclude the free evolution $R_\omega$ leaving only $R$.
Denoting $\Omega(t)=-d\phi(t)/dt$, for $\epsilon=0$, the evolution $U(t)=\mathcal T\exp\int^t H'(t')dt'/i$ is equivalent to (\ref{gat})
with $\epsilon=\varepsilon e^{-2i\theta}$ so the correction  reads
\be
\delta p_{1\theta}=\sin\theta(\sin 2\theta\mathrm{Re}\varepsilon+\cos 2\theta\mathrm{Im}\varepsilon)/2
\ee
Note the absence of 2nd harmonics so this model is insufficient to explain the found deviations.

\section*{Abbreviations}
RWA, rotation wave approximation; I/Q, in-phase/quadrature.
\section*{Declarations}
\subsection*{Ethical Approval and Consent to participate}
Not applicable.
\subsection*{Consent for publication}
No applicable.
\subsection*{Availability of supporting data}
The data are publicly available at https://zenodo.org/record/7538941
\subsection*{Competing interests}
The authors declare no competing interests.
\subsection*{Funding}
TR acknowledges the financial support by TEAM-NET project co-financed by EU within the Smart Growth Operational Programme 
(contract no.  POIR.04.04.00-00-17C1/18-00.
\subsection*{Authors' contributions}
T.B. collected the data and analyzed them. T.B. and T.R. wrote the scripts. J.T. and A.B. wrote the manuscript.
All the Authors reviewed the manuscript.

\end{document}